\documentclass[sigconf]{acmart} 
\usepackage[utf8]{inputenc} 
\usepackage{hyperref}       
\usepackage{url}            
\usepackage{amsfonts}       
\usepackage{nicefrac}       
\usepackage{amsthm}
\usepackage{enumitem}
\usepackage{xspace}
\usepackage{colortbl}
\usepackage{graphics}
\usepackage{multirow}
\usepackage{makecell}
\usepackage{graphicx}
\usepackage{subcaption}
\usepackage{algorithmic}

\usepackage[T1]{fontenc}    
\usepackage{framed}
\usepackage{booktabs}       
\usepackage{microtype}      
\usepackage{xcolor}         

\AtBeginDocument{%
  }

\copyrightyear{2026}
\acmYear{2026}
\setcopyright{cc}
\setcctype{by}
\acmConference[RecSys '26]{20th ACM Conference on Recommender Systems}{September 27-October 02, 2026}{Minneapolis, MN, USA}
\acmBooktitle{20th ACM Conference on Recommender Systems (RecSys '26), September 27-October 02, 2026, Minneapolis, MN, USA}
\acmDOI{10.1145/3773078.3831753}
\acmISBN{979-8-4007-2284-4/2026/09}

\definecolor{fgreen}{rgb}{0.0,0.5,0.0}

\newcommand{\raremed}{rare-med\xspace}
\newcommand{\raremeds}{rare-meds\xspace}
\newcommand{\freqmeds}{freq-meds\xspace}

\newcommand{\code}{\url{https://github.com/ShinhwanKang/GenRxR}}
\newcommand{\method}{\textsc{GenRxR}\xspace}

\newcommand{\numbaselines}{14\xspace}
\newcommand{\improvement}{30.9}

\newcommand\cforange[1]{\textcolor{orange}{#1}}   
\newcommand\cfpurple[1]{\textcolor{purple}{#1}}   
\newcommand\cfolive[1]{\textcolor{olive}{#1}}     
\newcommand\cfcyan[1]{\textcolor{cyan}{#1}}      
\newcommand\cfteal[1]{\textcolor{teal}{#1}}       
\newcommand\cfbrown[1]{\textcolor{brown}{#1}}     

\newcommand{\E}{\mathcal{E}}
\newcommand{\R}{\mathcal{R}}

\newcommand{\A}{\mathcal{A}}

\newcommand{\inputtextcf}{\mathcal{I}_{\text{CF}}}
\newcommand{\outputtextcf}{\mathcal{O}_{\text{CF}}}

\newcommand{\inputtextinst}{\mathcal{I}_{\text{IT}}}

\newcommand{\inputtextrec}{\mathcal{I}_{\text{rec}}}

\newcommand{\gentextsummary}{\hat{\mathcal{O}}^{i}_{\text{IT}}}
\newcommand{\textsummary}{\mathcal{O}^{i}_{\text{IT}}}

\newcommand{\D}{\mathcal{D}}
\newcommand{\Proc}{\mathcal{P}}
\newcommand{\M}{\mathcal{M}}

\newcommand{\targetmedi}{m^{*}}
\newcommand{\assoelements}{\mathcal{C}_{\targetmedi}}

\newcommand{\llmcf}{\mathbf{LLM}_{\mathbf{CF}}} 
\newcommand{\llminst}{\mathbf{LLM}_{\mathbf{IT}}} 
\newcommand{\llmrec}{\mathbf{LLM}_{\mathbf{REC}}} 
\newcommand{\loraparams}{\theta_{\text{LoRA}}}

\newtheoremstyle{problemstyle}  
{3pt}                                               
{3pt}                                               
{\normalfont}                               
{}                                                  
{\bfseries\itshape}                 
{\normalfont\bfseries:}         
{.5em}                                          
{}                                                  
\theoremstyle{problemstyle}

\newcommand{\smallsection}[1]{{\noindent {\bf{\underline{\smash{#1}}}}}}
\newcommand{\uls}[1]{\underline{\smash{#1}}}

\newcommand{\mimicthree}{MIMIC-\uppercase\expandafter{\romannumeral3}\xspace}
\newcommand{\mimicfour}{MIMIC-\uppercase\expandafter{\romannumeral4}\xspace}

\newcommand{\jaccard}{\mathrm{Jaccard}}
\newcommand{\fone}{\mathrm{F1}}

\definecolor{strongyellow}{RGB}{255,255,0}
\definecolor{orangehighlight}{RGB}{255,200,0}  
\definecolor{limehighlight}{RGB}{200,255,0}
\definecolor{lightgray}{RGB}{220,220,220}    
\newcommand{\best}[1]{\textbf{\colorbox{strongyellow}{#1}}}
\newcommand{\secbest}[1]{\underline{#1}}
\newcommand{\thibest}[1]{#1}

\newcommand{\israre}[1]{\textbf{\colorbox{orangehighlight}{#1}}}
\newcommand{\allright}[1]{\textbf{\colorbox{lightgray}{#1}}}

\newcommand{\checkimg}{\strut\includegraphics[scale=0.02]{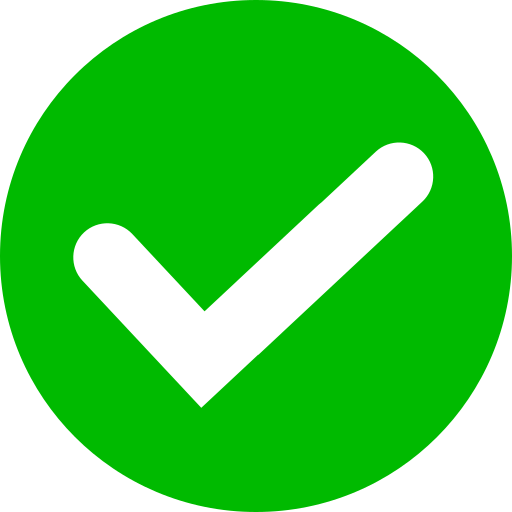}}

\begin{document}

\title[Improving Rare Medication Recommendation]{Improving Rare Medication Recommendation with Counterfactual Data Augmentation and Large Language Models}
\settopmatter{authorsperrow=3}

\author{Shinhwan Kang}
\email{shinhwan.kang@kaist.ac.kr}
\affiliation{%
  \institution{KAIST}
  \city{Seoul}
  \country{South Korea}
}
\author{Soo Yong Lee}
\email{syleetolow@kaist.ac.kr}
\affiliation{%
  \institution{KAIST}
  \city{Seoul}
  \country{South Korea}
}
\author{Jaewon Kim}
\email{jwk1921@kaist.ac.kr}
\affiliation{%
  \institution{KAIST}
  \city{Daejeon}
  \country{South Korea}
}
\author{Kijung Shin}
\authornote{Co-corresponding authors.} 
\email{kijungs@kaist.ac.kr}
\affiliation{%
  \institution{KAIST}
  \city{Seoul}
  \country{South Korea}
}
\author{Buru Chang}
\authornotemark[1]
\email{buru_chang@korea.ac.kr}
\affiliation{%
  \institution{Korea University}
  \city{Seoul}
  \country{South Korea}
}  

\renewcommand{\shortauthors}{Kang et al.}

\begin{abstract}
AI-based medication recommendation systems have attracted substantial attention due to their potential to enhance patient safety and therapeutic outcomes.
Despite the clinical importance of accurately recommending rarely prescribed medications (\raremeds), we observe that most existing methods show significantly lower predictive performance for \raremeds.
We attribute this issue to two intrinsic limitations: 
(a) the inherent scarcity of data for \raremeds and 
(b) limited consideration of co-recommended medications.
To address these limitations, we propose \method, a novel framework based on large language models (LLMs). 
\method leverages the medical knowledge and clinical reasoning capability of LLMs to generate counterfactual medical data, mitigating the data scarcity issue for \raremeds.
It also integrates an LLM into the medication recommendation process to model relationships among co-recommended medications.
To further enhance the clinical reasoning, we introduce an instruction tuning step that aligns the LLM's capability with the recommendation task, enabling better handling of clinical context, including \raremeds cases.
In our experiments, we show that \method outperforms \numbaselines (including 5 LLM-based) baselines in most cases.
Specifically, it achieves up to \improvement\% higher predictive performance for \raremeds than the strongest baseline.
\end{abstract}

%
%


%



\maketitle

\vspace{-9pt}
\section{Introduction}
\label{sec:intro}

Prescribing appropriate medications is essential for ensuring patient safety and achieving effective therapeutic outcomes. 
To support decision-making in medication prescription,
artificial intelligence (AI)-based medication recommendation systems~\citep{shang2019gamenet,yang2021safedrug,zhao2024leave,tan2024natural,wu2022conditional,yang2023molerec} have recently emerged as a promising research direction. 
These AI-based approaches are expected to support personalized treatment by adapting recommendations to individual patient characteristics,
with the potential to perform comparably to human experts.

\begin{figure}[!t]
    \centering
    \includegraphics[width=\linewidth]{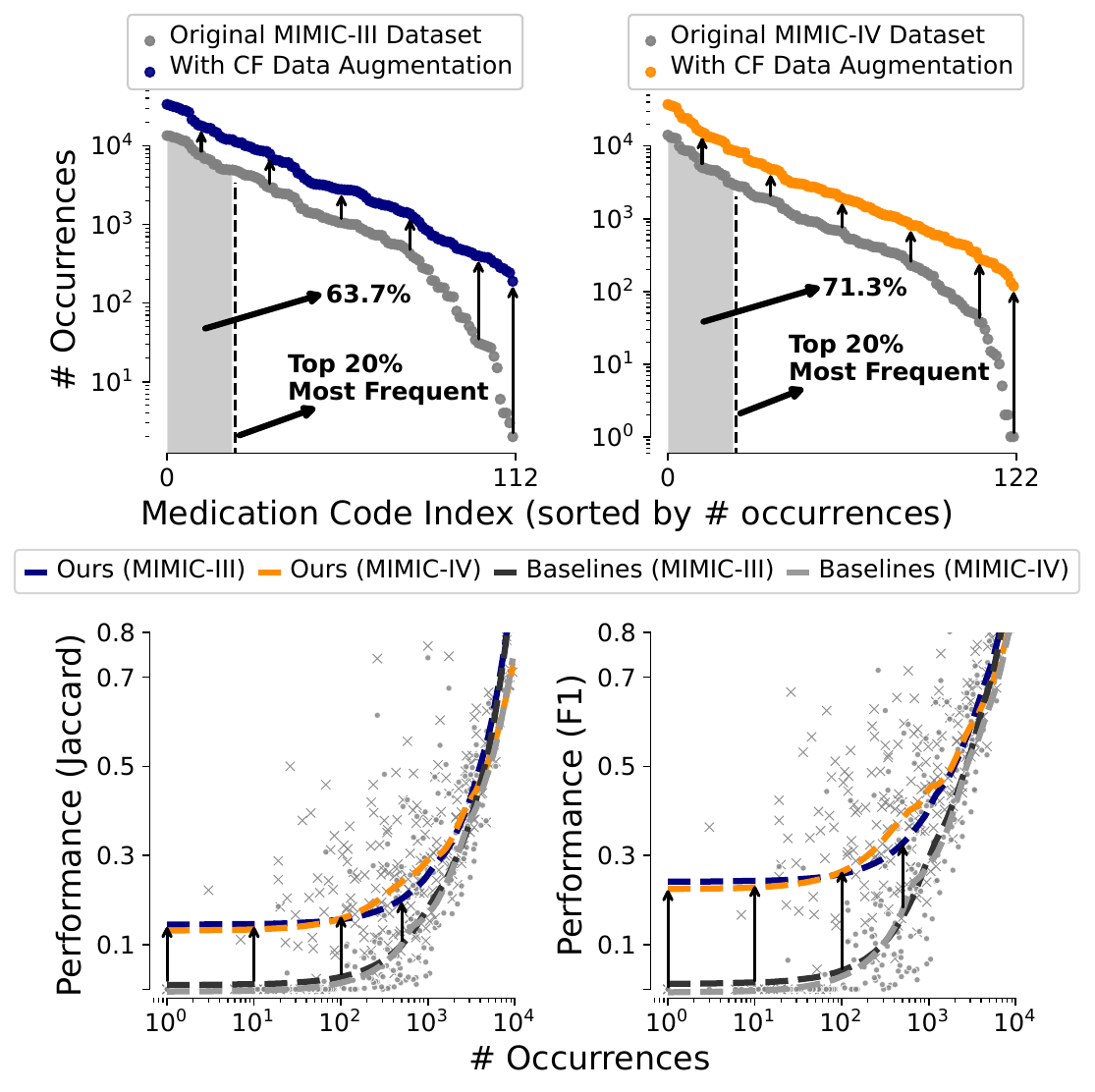}
    \caption{(a) Medication Occurrence Frequency (Top) and (b) Average Predictive Performance across Frequencies (Bottom). 
    Medications in original EHR datasets exhibit long-tail distributions, with many rarely prescribed medications (\raremeds), and predictive performance drops significantly for \raremeds.
    Our approach, including counterfactual (CF) data augmentation, effectively mitigates this issue.}
    \label{fig:crown_figure}
    \Description[(a) Medication Occurrence Frequency (Top) and (b) Average Predictive Performance across Frequencies (Bottom).]{The upper figure describes Medication Occurrence Frequency and the lower figure describes Average Predictive Performance across Frequencies. 
    Medications in original EHR datasets exhibit long-tail distributions, with many rarely prescribed medications (\raremeds), and predictive performance drops significantly for \raremeds.
    Our approach, including counterfactual (CF) data augmentation, effectively mitigates this issue.}
\end{figure}

Despite their promise for personalized and accurate prescriptions, our analysis reveals that existing AI-based methods exhibit significant prediction biases related to prescription frequency.
Specifically, as illustrated in Figure~\ref{fig:crown_figure}(b), most existing methods~\citep{yang2021safedrug,choi2016retain,shang2019gamenet,wu2022conditional} show concerningly poor predictive performance on \textit{rarely prescribed medications} (\raremeds) compared to the frequent counterparts, often completely failing to predict \raremeds.\footnote{The lower predictive performance for \raremeds resembles the item cold-start problem, as prescription instances involving \raremeds are limited in the training data. Yet medication recommendation must also account for clinical consistency and safety constraints, which distinguishes our setting from typical item cold-start settings.}

The inability to accurately recommend \raremeds significantly limits the clinical utility of current systems.
Many \raremeds, such as orphan drugs~\citep{nih_orphan}, are specifically designed to treat uncommon or severe medical conditions that typically have limited therapeutic options~\citep{yoo2023development,ema_orphan,fecho2022leveraging}. 
Consequently, failure to accurately prescribe these medications can endanger patient safety and lead to severe or life-threatening outcomes.

\textit{Why do existing AI-based medication recommendation systems exhibit substantially lower predictive performance on \raremeds?}
The primary reason is the inherent data scarcity associated with \raremeds.
Due to strict privacy regulations and ethical considerations, electronic health record (EHR) datasets are rarely available and contain especially few instances of \raremed prescriptions.
Indeed, as illustrated in Figure~\ref{fig:crown_figure}(a), medications exhibit a long-tail distribution, where a small subset of medication types accounts for the majority of occurrences, 
whereas the others, including \raremeds, appear significantly less frequently.
Due to their data-driven nature, AI-based models are strongly affected by this imbalance, leading to low predictive performance for \raremeds. 
Another important factor is that methods exhibiting the performance pattern in Figure~\ref{fig:crown_figure}(b) typically pay limited attention to the relationships among co-recommended medications, which can provide critical clinical context for \raremeds prescription.

To address the aforementioned limitations, we propose a novel medication recommendation framework, \textbf{\method}.
It leverages large language models (LLMs), which are known to have extensive medical knowledge and clinical reasoning capabilities~\citep{singhal2025toward,kungperformance,nori2023capabilities}, for (1) counterfactual data augmentation and (2) medication recommendation, where consideration of clinical context is crucial.

Specifically, we introduce an LLM-based counterfactual data augmentation method to alleviate data scarcity. 
It generates hypothetical (`what-if') scenarios in which a patient receives medications that were not actually prescribed to them.
These scenarios are provided to the LLM, which then infers clinical changes in diagnoses, procedures, and co-prescribed medications to produce synthetic patient data.
This enriches training datasets and enables our recommendation model to learn more diverse clinical patterns.

By integrating an LLM into our medication recommendation model, \method naturally considers information of co-recommended medications when recommending subsequent ones.
To further enhance the capability and better align the LLM's clinical reasoning with the medication recommendation task, we introduce instruction tuning involving medical record summarization relevant to a specific medication. 
This step enhances the capture of patient's clinical context, including cases where \raremeds need to be recommended.

As a result, as shown in Figure~\ref{fig:crown_figure}, (a) the generated counterfactual data alleviates data scarcity, particularly mitigating the lack of \raremeds in the training set, and (b) the overall \method framework enhances predictive performance for \raremeds without compromising performance on frequently prescribed medications.
Numerically, \method achieves up to \textbf{\improvement\% better predictive performance} for \raremeds compared to existing \numbaselines baselines methods.

Our key contributions are summarized as follows:
\vspace{-2pt}
\begin{itemize}[leftmargin=*]
    \item \textbf{Novel and Crucial Focus:} We reveal a critical yet overlooked limitation of existing AI-based medication recommendation methods in handling \raremeds and propose mitigation strategies.
    \item \textbf{Counterfactual Medical Data}: To the best of our knowledge, we propose the first counterfactual data augmentation strategy for medication recommendation. To this end, we leverage the medical knowledge and clinical reasoning capability of LLMs.
    \item \textbf{Effective Usage of LLMs}: By leveraging an instruction-tuned LLM, \method significantly outperforms four LLM-based baselines and nine state-of-the-art medication recommender systems.
\end{itemize}
For \textbf{reproducibility}, we release the code and generated counterfactual data at \textit{\code}.
\vspace{-5pt}

\section{Related Work}
\label{sec:related}

\smallsection{Medication Recommendation.}
A variety of medication recommendes have been developed, with most leveraging the electronic health records (EHRs) of patients as input.
RETAIN~\citep{choi2016retain} employs a two-level neural attention mechanism to identify influential past visits and significant clinical variables within those visits.
GAMENet \citep{shang2019gamenet} uses drug–drug interaction (DDI) graphs and medication co-occurrence graphs to derive informative medication embeddings, subsequently utilizing a memory network to incorporate patient-specific historical data.
SafeDrug~\citep{yang2021safedrug} leverages molecular information via message-passing neural networks to construct chemically informed medication embeddings.
MoleRec~\citep{yang2023molerec} further enhances medication representation by explicitly modeling molecular substructures, motivated by the hypothesis that specific substructures correlate strongly with particular diseases.
RAREMed~\citep{zhao2024leave} introduces self-supervised pre-training strategies to strengthen embeddings of diagnoses and procedures, addressing patient-level fairness concerns.
LEADER~\citep{liu2024large} incorporates an LLM by adding a classifier to its last hidden states, simultaneously predicting all medications at once. 
LAMO~\citep{zhao2025fine} and FLAME~\citep{fan2025fine} leverage discharge summaries using LLMs and make binary prescription decisions over all candidate medications, after which FLAME further refines the initial recommendation set through a two-stage framework.
Notably, a few studies have formulated medication recommendation as a sequential decision-making problem.
For instance, 
LEAP~\citep{zhang2017leap} uses a recurrent decoder to predict medications sequentially, and 
COGNet~\citep{wu2022conditional} utilizes an iterative transformer decoder that recommends a medication based on the previously predicted medications and the patient information.
However, most existing methods perform poorly on \raremeds and lack dedicated mechanisms to address this challenge.

\smallsection{Data Augmentation using LLMs.}
Data augmentation~\citep{dang2025data,ghosh2023bioaug,dai2022promptagator} has been widely adopted to address data scarcity across various machine learning domains, including the medical domain~\citep{wiese2021nncompare,fan2020relational,amad2025improving}, where such scarcity is particularly severe.
Recently, LLMs have shown promise in synthesizing medical data~\citep{ashofteh2025targeted,liu2025generating,smolyak2024large}. 
For example, GatorTronGPT~\citep{yang2022large} is fine-tuned to generate clinical notes, leading to improved performance in clinical NLP tasks~\citep{peng2023study}.
LLMs have also been used to generate counterfactual samples, mostly in general-domain settings~\citep{sen2023people,li2023prompting,ravfogelgumbel,balashankar2023improving}.
For instance, DISCO~\citep{chen2022disco} prompts LLMs to minimally modify input text under counterfactual label conditions, improving robustness in inference.
However, to the best of our knowledge, prior works have not specifically addressed counterfactual augmentation for medication recommendation, which is a focus of this work.

\smallsection{Recommender Systems with LLMs.}
With the advancement of LLMs, many studies have integrated them into various recommendation tasks~\citep{bao2023tallrec,cui2022m6,dai2023uncovering,geng2022recommendation,hou2024large,wu2024survey} to leverage textual information beyond traditional ID-based representations.
However, recent works~\citep{zhang2025collm,liao2024llara} pointed out that relying solely on text features may overlook collaborative patterns and proposed methods that jointly leverage ID-based and text-based representations to combine collaborative signals with textual semantics.
\section{Preliminaries and Problem Definition}
\label{sec:preliminaries}

In this section, we define the Electronic Health Record (EHR) data and the medication recommendation task.

\smallsection{Data.} 
A hospital EHR database is a comprehensive digital repository that stores various aspects of patient information, including \textit{diagnoses}, \textit{procedures}, and prescribed \textit{medications}.
In this study, we utilize EHR datasets (\mimicthree~\citep{johnson2016mimic} and \mimicfour~\citep{johnson2023mimic}) derived from such databases for research purposes.
Let $\E = \{\R^{i}\}_{i=1}^{N}$ denote the structured data, where $N$ represents the total number of patients. 
We assume each patient's record $\R^{i}$ can be described as a sequence of hospital admissions, denoted as $[\A^{i}_1, \A^{i}_2, \dots, \A^{i}_{T_{i}}]$, where $T_{i}$ is the total number of admissions for patient $i$. 
Each admission $\A^{i}_{t}$ at time $t$ consists of three primary components: the set of diagnoses $\D^{i}_{t}$, the set of medical procedures performed $\Proc^{i}_{t}$, and the set of prescribed medications $\M^{i}_{t}$. 
Formally, the $t$-th admission of patient $i$ can be expressed as $\A^{i}_{t} = \{ \D^{i}_{t}, \Proc^{i}_{t}, \M^{i}_{t} \}$.

\smallsection{Problem Definition.}
Medication recommendation~\citep{choi2016retain,zhang2017leap,yang2021safedrug,yang2021change,wu2022conditional,yang2023molerec,zhao2024leave,liu2024large} aims to predict the medications that a patient should be prescribed at a given time $t$ based on their current admission information and medical history.  
Formally, given a patient's record $\R^{i}$, excluding the current prescription $\M^{i}_{t}$, the objective is 
to produce a predicted medication set $\hat{\M}^{i}_{t}$ closely matching the true set $\M^{i}_{t}$.
\section{\method: Proposed Framework for Rare-Meds}
\label{sec:proposed_framework}
In this section, we propose \method, a framework that leverages two large language models (LLMs) to enhance recommendations for rarely prescribed medications (\raremeds).

\subsection{Counterfactual Data Generation with LLMs}
\label{sec:cf_gen}
To mitigate the data scarcity issue associated with rarely prescribed medications (\raremeds), we propose a counterfactual data generation method leveraging an LLM. 
Since recent LLMs have been reported to possess strong clinical~\citep{singhal2025toward,kungperformance,nori2023capabilities} and counterfactual reasoning capacity~\citep{chen2022disco,li2023prompting}, we expect them to generate high-quality counterfactual clinical data. 
The goal of this step is to systematically create contextually plausible and hypothetical patient data reflecting diverse clinical patterns, particularly for \raremeds.

\begin{figure}[t]
    \centering
    \includegraphics[width=1.0\linewidth]{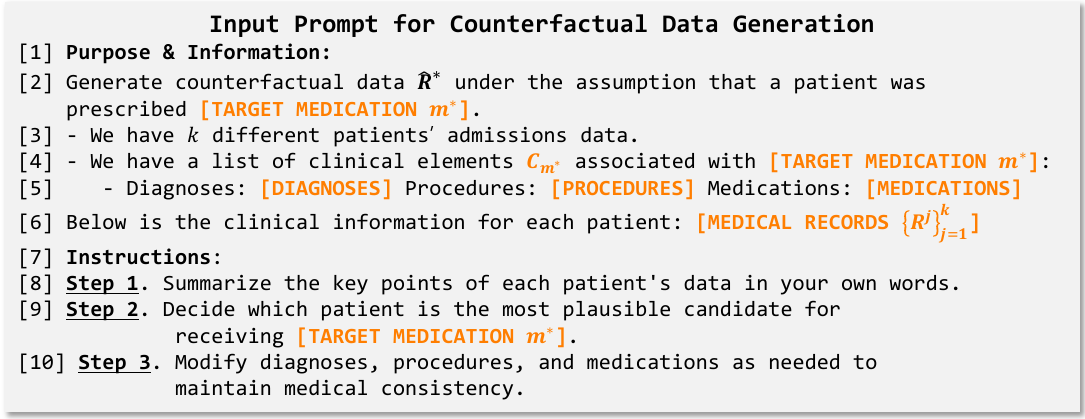}
    \caption{Input Prompts for Counterfactual Data Generation.}
    \label{fig:cf_prompt}
    \Description[Input Prompts for Counterfactual Data Generation]{This figure describes an input prompt for counterfactual data generation.}
\end{figure}

Our counterfactual data generation method, whose prompt is outlined in Figure~\ref{fig:cf_prompt}, consists of two steps: (1) \textit{counterfactual scenario construction}, where we define hypothetical ``what-if'' scenarios in which a patient is assumed to receive a medication that has not been prescribed, and (2) \textit{counterfactual inference}, where an LLM infers plausible changes in diagnoses, procedures, and co-prescribed medications under such scenarios.
We describe each step below.

\smallsection{Counterfactual Scenario Construction.}
We first construct hypothetical counterfactual scenarios by assuming the introduction of a specific medication into the clinical history of a patient who has not previously been prescribed that medication. Specifically, given a patient $i$'s medical record $\R^{i} = [\A^{i}_1, \dots, \A^{i}_{T_i}]$, we define the aggregated set of medications prescribed throughout the patient's admissions as $\M^{i} = \bigcup_{t=1}^{T_i}\M^{i}_{t}$. 
Then, we formulate a counterfactual scenario by hypothetically prescribing a target medication $m^{*}$ to patient $i$ who has no historical prescription record of $m^{*}$ as follows:
\[
\M^{i*} \leftarrow \M^{i} \cup \{m^{*}\}, \text{ where } m^{*} \notin \M^{i}.
\]
This hypothetical change in the set of prescribed medications $\M^{i}$ leads to the following counterfactual question:
\vspace{-3pt}
\begin{framed}
\textit{If medication $m^{*}$ had been prescribed to patient $i$, what contextually plausible changes in diagnoses, procedures, and medications would occur in the patient's medical record\xspace$\R^{i}$?}
\end{framed}
\vspace{-3pt}

\begin{figure}[t]
    \vspace{-1mm}
    \centering
    \includegraphics[width=1.0\linewidth]{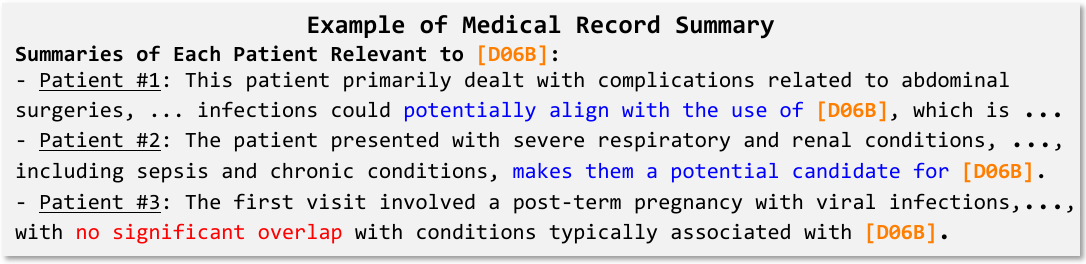}
    \caption{Example of Medical Record Summary.}
    \label{fig:medical_summary}
    \Description[Example of Medical Record Summary]{This figure describes an example of medical record summary.}
\end{figure}

\smallsection{Counterfactual Inference Using an LLM.}
To answer this counterfactual question, we employ an LLM, denoted as $\llmcf(\cdot)$. 
This LLM generates counterfactual data by explicitly reasoning about plausible changes to the patient's clinical data in the counterfactual scenario. 
Specifically, the inference involves reasoning steps that include selecting a clinically appropriate patient among $k$ randomly-chosen candidates and subsequently generating the corresponding counterfactual clinical data, as shown in Figure~\ref{fig:cf_prompt}.
Note that using an LLM to select the most plausible candidate for prescribing
$\targetmedi$ (Lines~8--9 of Figure~\ref{fig:cf_prompt}) helps avoid clinically inconsistent scenarios, where its prescription would be inappropriate.
Formally, this counterfactual inference procedure is:
\[
\outputtextcf, \hat{\R}^{*} = \llmcf(\inputtextcf,\, \{\R^{j}\}_{j=1}^{k},\, m^{*},\assoelements).
\]
\noindent Details of inputs and outputs are summarized below.
\begin{framed}
\smallsection{Inputs.}
Inputs given to $\llmcf$ (see Figure~\ref{fig:cf_prompt}) consist of:
\begin{itemize}[leftmargin=*, topsep=1pt, itemsep=1pt]
\item $\inputtextcf$: Instruction to guide the counterfactual inference.
\item $\{\R^{j}\}_{j=1}^{k}$: Medical records of $k$ candidate patients.
\item $m^{*}$: Target medication hypothetically added.
\item $\assoelements$: Set of clinical elements (i.e., diagnosis, procedure, and medication) associated with the target medication $\targetmedi$.
\end{itemize}
\end{framed}
\vspace{-3pt}
\begin{framed}
\smallsection{Outputs.}
The $\llmcf$ generates the following outputs:
\begin{itemize}[leftmargin=*, topsep=1pt, itemsep=1pt]
\item $\outputtextcf$: Generated text consisting of 
(1) summaries of candidate patients' clinical contexts (\textit{medical record summary}, see Figure~\ref{fig:medical_summary}) and (2) explicit rationale for selecting the most suitable patient for prescribing the target medication $\targetmedi$.
\item $\hat{\R}^{*}$: Generated counterfactual data for the selected patient, including the modified diagnoses, procedures, and medications.
\end{itemize}
\end{framed}
\vspace{-3pt}
\noindent 
As indicated in the inputs above, the associated clinical elements (i.e., diagnoses, procedures, and medications) $\assoelements$ are provided to $\llmcf$ to promote contextually coherent counterfactual data generation (refer to Lines~4--5 of Figure~\ref{fig:cf_prompt}).
Specifically, these associated elements are selected based on their statistical association with the target medication $\targetmedi$, quantified by \textit{relative risk} as follows\footnote{We use relative risk as a contextual signal for selecting associated clinical elements $\assoelements$.
Although relative risk may be unreliable for medications observed in very few samples, we reduce this risk by instructing $\llmcf$ to ignore irrelevant elements, select the most plausible patient, and preserve clinical consistency.}:
\[
\mathrm{RelativeRisk}(c, \targetmedi)
= \frac{P(c \mid \targetmedi)}{P(c \mid \neg \targetmedi)}
= \frac{\mathrm{count}(c, \targetmedi) / \mathrm{count}(\targetmedi)}
       {\mathrm{count}(c, \neg \targetmedi) / \mathrm{count}(\neg \targetmedi)},
\]
where $\text{count}(c, \targetmedi)$ denotes the number of admissions containing both the clinical element $c$ and the target medication $\targetmedi$, and $\neg \targetmedi$ denotes admissions without $\targetmedi$.  
A high relative risk indicates that a clinical element frequently co-occurs with the target medication $\targetmedi$ but is uncommon otherwise, making it a strong candidate for inclusion in $\assoelements$ to support plausible counterfactual generation.

As described in the outputs above, $\llmcf$ generates comprehensive summaries of candidate patients’ clinical contexts, along with the rationale for selecting the most clinically suitable patient for prescribing the target medication $\targetmedi$
(Lines 8--9 of Figure~\ref{fig:cf_prompt}; see Figure~\ref{fig:medical_summary} for an example summary).
These summaries and rationales provide the contextual foundation when $\llmcf$ subsequently generates counterfactual data $\hat{\R}^{*}$ (Line 10 of Figure~\ref{fig:cf_prompt}; see Table~\ref{tab:anal_cf_data} for examples of generated counterfactual data).
Note that the generated summary is also used in Section~\ref{sec:inst_tune} for instruction tuning of another LLM that performs medication recommendation, and that our augmentation method can be applied repeatedly.

\subsection{Instruction Tuning for Enhancing Clinical Reasoning Capability of LLMs}
\label{sec:inst_tune}
In this subsection, we present the instruction tuning of an LLM through medical record summarization. 
To this end, we use summaries generated as described in the previous subsection by using another LLM with stronger medical interpretation capabilities. 
Note that this step precedes further LLM fine-tuning for medication recommendation, which is described in the next subsection.

Instruction tuning refers to a fine-tuning process that equips an LLM with broadly applicable capabilities (in our context, clinical reasoning) before it is further fine-tuned for specific tasks (in our context, medication recommendation).
For instruction tuning in our setting, we employ a medical record summarization task where the LLM summarizes a patient’s medical records with a focus on information relevant to prescribing a specific medication.
Through this process, we expect the LLM to enhance its ability to interpret patient records and assess medication-relevant clinical information, ultimately enabling it to make informed prescription decisions.

\begin{figure}[t]
    \vspace{-1mm}
    \centering
    \includegraphics[width=1.0\linewidth]{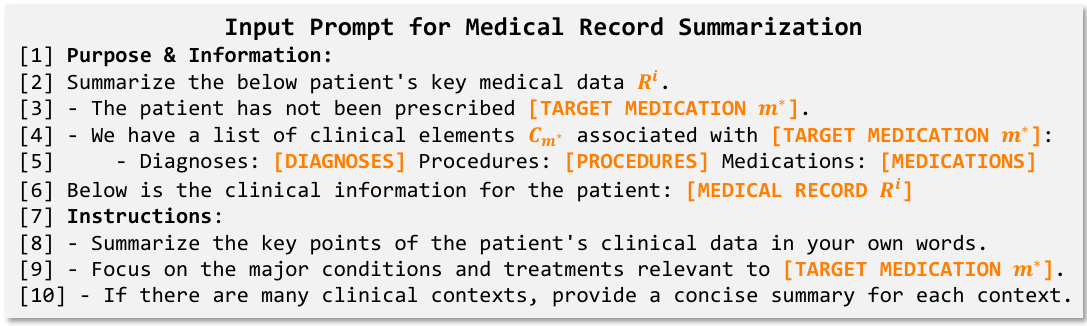}
    \caption{Input Prompts for Instruction Tuning based on Medical Record Summarization.}
    \label{fig:it_prompt}
    \Description[Input Prompts for Instruction Tuning based on Medical Record Summarization]{This figure describes an input prompt for instruction tuning based on medical record summarization.}
\end{figure}

The prompt for instruction tuning is outlined in Figure~\ref{fig:it_prompt}.
Given an instruction $\inputtextinst$, a patient medical record $\R^{i}$, a target medication $\targetmedi$, and associated clinical elements $\assoelements$, a fine-tunable LLM denoted by $\llminst$ generates a medical record summary $\hat{\mathcal{O}}^{i}_{IT}$, i.e.,
\[
\gentextsummary = \llminst(\inputtextinst, \R^{i}, \targetmedi, \assoelements),
\]
where the instruction $\inputtextinst$ guides the model to summarize key clinical elements relevant to prescribing medication $\targetmedi$ (refer to Lines 8--10 of Figure~\ref{fig:it_prompt}). 
$\llminst$ is fine-tuned to generate an output $\gentextsummary$ that captures the patient’s major clinical conditions and treatments relevant to prescribing the medication $\targetmedi$ (see Figure~\ref{fig:medical_summary} for an example), as described in detail below.

\begin{figure*}[!ht]
    \vspace{-1mm}
    \centering
    \includegraphics[width=1.0\linewidth]{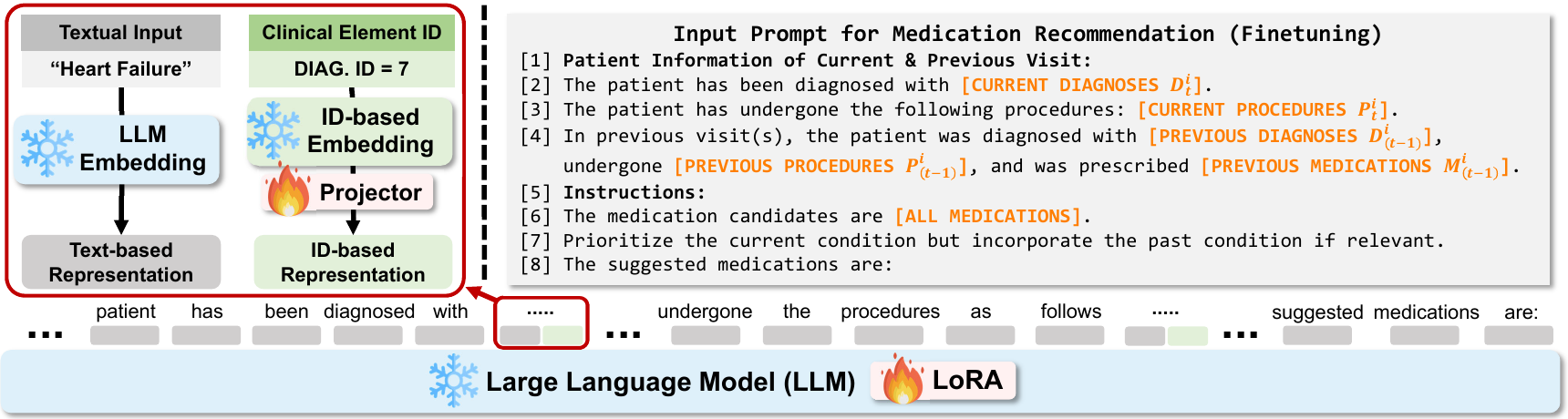}
    \caption{Overview of LLM-based Medication Recommendation Model $\llmrec$.}
    \label{fig:overview_model}
    \Description[Overview of LLM-based Medication Recommendation Model]{This figure describes an overview of LLM-based medication recommendation model.}
\end{figure*}

\smallsection{Training.}
We optimize $\llminst$ using parameter-efficient adaptation techniques (LoRA)~\citep{hu2022lora} using
patient summaries generated by $\llmcf$ in the counterfactual data generation step (Section~\ref{sec:cf_gen}) as ground-truth summaries.
Note that $\llminst$, which is fine-tunable, is distinct from $\llmcf$, which is not necessarily fine-tunable but is expected to possess advanced clinical understanding (e.g., proprietary LLMs).
Specifically, the training objective is defined as:
\[
\mathcal{L}_{\mathbf{IT}} = - \sum\nolimits_{j=1}^{|\textsummary|}\log p(\mathcal{O}^{i}_{\text{IT},j} \mid \inputtextinst, \mathcal{R}^{i}, \targetmedi, \assoelements, \mathcal{O}^{i}_{\text{IT},<j}; \loraparams),
\]
where $\textsummary$ denotes the ground-truth summary of the medical record for patient $i$, 
$\mathcal{O}^{i}_{\text{IT},j}$ is the $j$-th token in $\textsummary$,
$\mathcal{O}^{i}_{\text{IT},<j}$ is the sequence of previous $(j-1)$ tokens, and
$\loraparams$ is trainable LoRA parameters.

\subsection{Medication Recommendation with LLMs}
\label{sec:finetune}
In this subsection, we present our LLM-based medication recommendation model, denoted as $\llmrec$. We obtain $\llmrec$ by further fine-tuning $\llminst$, which is instruction-tuned in Section~\ref{sec:inst_tune}, and to this end, we use counterfactual data generated in Section~\ref{sec:cf_gen}.
The key idea of $\llmrec$ is to recommend medications one at a time, using previously recommended medications as clinical context, which can promote the (proper) recommendation of \raremeds. 

Our medication recommendation model,  $\llmrec$, is outlined in Figure~\ref{fig:overview_model}.
$\llmrec$ takes an instruction $\inputtextrec$ (Lines 5--8 of Figure~\ref{fig:overview_model}) and patient clinical data as inputs.
The instruction $\inputtextrec$ guides $\llmrec$ to recommend medications based on the given patient's clinical data (Lines 2--4 of Figure~\ref{fig:overview_model}).
Patient clinical data consists of current diagnoses $\D^{i}_{t}$, procedures $\Proc^{i}_{t}$, and admission histories $\mathcal{R}^{i}_{<t}$, the latter available only for patients with prior visits. 
Formally, $\llmrec$ generates the next token as follows:
\[
\hat{\mathbb{M}}^{i}_{t, j} = \llmrec(\inputtextrec,\, \mathcal{R}^{i}_{<t}, \mathcal{D}^{i}_{t},\, \mathcal{P}^{i}_{t}, \hat{\mathbb{M}}^{i}_{t, <j}),
\]
where $\hat{\mathbb{M}}^{i}_{t,j}$ denotes the $j$-th predicted token for patient $i$ at $t$-th admission, 
$\mathcal{R}^{i}_{<t}$ denotes admission histories, and 
$\hat{\mathbb{M}}^{i}_{t, <j}$ is the sequence of previously generated $(j-1)$ tokens (i.e., previously recommended medications).
Note that by taking previously recommended medications as input, $\llmrec$ captures their interrelationships and establishes a clinical context for subsequent recommendations.
We next specify how $\llmrec$ represents clinical elements used above.

\smallsection{Representations of Clinical Elements.}
As illustrated on the left side of Figure~\ref{fig:overview_model}, each clinical element is represented in two ways. 
First, a text-based representation is obtained by encoding the textual title (e.g., ``Heart Failure'') with the LLM embedding. 
Second, an ID-based representation is obtained by passing the element’s unique-ID-based initial embedding through a projection layer that maps it into the same latent space as the text embedding.
Then, these two representations are provided jointly to $\llmrec$ as separate tokens in the input sequence.
The text-based representation conveys semantic meaning, whereas the ID-based representation is expected to capture association patterns inherent in the training data, consistent with recent LLM-based recommendation findings~\citep{zhang2025collm,liao2024llara}.

\smallsection{Training \& Inference.}
Initialized with the parameters of $\llminst$ (Section~\ref{sec:inst_tune}), we fine-tune the entire set of trainable parameters of $\llmrec$, including auxiliary parameters (e.g., a projection layer), 
using the original EHR dataset together with the augmented data generated by our counterfactual approach (Section~\ref{sec:cf_gen}).
To this end, we aim to maximize the likelihood of predicting the next token in the medication sequence derived from the ground-truth medication sets.
\footnote{We convert each ground-truth medication set into a medication sequence ordered by ascending frequency in the training set, to prioritize rarely prescribed medications.}
The training objective of $\llmrec$ is defined as:
\[
\mathcal{L}_{\text{rec}} = - \sum\nolimits_{j=1}^{|\mathbb{M}^{i}_{t}|}\log p(\mathbb{M}^{i}_{t,j}\mid \inputtextrec,  \mathcal{R}^{i}_{<t}, \D^{i}_{t}, \Proc^{i}_{t}, \mathbb{M}^{i}_{t, <j}; \theta),
\]
where 
$\mathbb{M}^{i}_{t}$ is the textual token sequence of ground-truth medication set for patient $i$ at $t$-th admission,
$\mathbb{M}^{i}_{t,j}$ denotes the $j$-th token in $\mathbb{M}^{i}_{t}$,
$\mathbb{M}^{i}_{t,<j}$ denotes the sequence of previous $(j-1)$ tokens in $\mathbb{M}^{i}_{t}$, and
$\theta$ represents all trainable parameters of $\llmrec$.
At inference time, we obtain the final predicted medication set $\hat{\mathcal{M}}^{i}_{t}$  by extracting medications from the generated textual sequence $\hat{\mathbb{M}}^{i}_{t}$.
\begin{table*}[!ht]
    \vspace{-2mm}
    \caption{Comparison of Medication Recommendation Methods on \mimicthree and \mimicfour. 
    The best result is highlighted in \best{yellow}, and the second-best result is \uls{underlined}. 
    Percentage values in parentheses under the results of \method denote relative improvements over the best-performing baseline for each metric.
    For DDI Ratio and the number of predicted medications, the 
    `GT' value in under the results of \method denotes the corresponding ground-truth value from the dataset.
    }
    \setlength{\tabcolsep}{1.0pt}
    \small
    \centering
    \scalebox{0.78}{
        \begin{tabular}{c|cccc|cc|cccc|cc|cc}
        \toprule
        \multirow{2}[3]{*}{\textbf{\mimicthree}} & \multicolumn{4}{c|}{\textbf{Group Jaccard}} & \multirow{2}[1]{*}{\textbf{\makecell{Micro\\Jaccard}}} & \multirow{2}[1]{*}{\textbf{\makecell{Macro\\Jaccard}}} & \multicolumn{4}{c|}{\textbf{Group F1}} & \multirow{2}[1]{*}{\textbf{\makecell{Micro\\F1}}} & \multirow{2}[1]{*}{\textbf{\makecell{Macro\\F1}}} & 
        \multirow{2}[1]{*}{\textbf{\makecell{DDI\\Ratio}}} & 
        \multirow{2}[1]{*}{\textbf{\makecell{\# Predicted\\Medications}}} \\
        \cmidrule(lr){2-5}\cmidrule(lr){8-11} 
        &  \textbf{1 (Rare)} & \textbf{2} & \textbf{3} & \textbf{4 (Freq.)} & & & \textbf{1 (Rare)} & \textbf{2} & \textbf{3} & \textbf{4 (Freq.)} & &  \\
        \midrule \midrule
\textbf{Logistic Reg.} & 2.08±0.00 & 14.12±0.00 & 26.46±0.00 & 54.76±0.00 & 49.05±0.00 & 24.36±0.00 & 3.50±0.00 & 23.25±0.00 & 39.59±0.00 & 68.60±0.00 & 64.90±0.00 & 33.74±0.00 & 7.73±0.00 & 17.06±0.00 \\
\textbf{RETAIN}~\citep{choi2016retain} & 1.07±1.40 & 11.39±1.38 & 24.15±0.32 & 56.00±0.31 & 50.39±0.08 & 23.21±0.30 & 1.88±1.94 & 18.57±2.05 & 35.95±0.53 & 69.65±0.48 & 66.15±0.08 & 31.62±0.39 & 7.46±0.09 & 25.47±0.13 \\
\textbf{LEAP}~\citep{zhang2017leap} & 1.29±0.27 & 6.76±0.91 & 12.94±0.45 & 44.73±0.23 & 42.99±0.21 & 16.43±0.34 & 2.35±0.46 & 11.84±1.40 & 20.56±0.74 & 57.63±0.29 & 59.29±0.24 & 23.10±0.60 & 7.27±0.10 & 13.85±0.12 \\
\textbf{GAMENet}~\citep{shang2019gamenet} & 2.42±0.57 & 9.44±0.43 & 23.56±0.41 & 55.74±0.11 & 50.99±0.15 & 22.77±0.32 & 3.76±0.82 & 15.48±0.74 & 34.86±0.58 & 69.16±0.12 & 66.61±0.15 & 30.82±0.51 & 7.69±0.15 & 19.08±0.17 \\
\textbf{SafeDrug}~\citep{yang2021safedrug} & 1.50±0.95 & 9.45±0.26 & 27.19±1.62 & 55.15±2.93 & 50.31±1.86 & 23.31±1.26 & 2.34±1.36 & 14.93±0.34 & 39.55±1.89 & 68.68±2.92 & 65.97±1.71 & 31.38±1.43 & 6.42±0.11 & 21.67±0.43 \\
\textbf{COGNet}~\citep{wu2022conditional} & \secbest{10.19±0.24} & \secbest{22.15±0.39} & \secbest{32.79±0.09} & \secbest{59.48±0.20} & 52.54±0.16 & \secbest{31.16±0.19} & \secbest{15.75±0.20} & \secbest{34.58±0.47} & \secbest{47.51±0.11} & \secbest{73.04±0.20} & 67.89±0.15 & \secbest{42.72±0.19} & 7.87±0.15 & 29.84±0.22 \\
\textbf{MICRON}~\citep{yang2021change} & 3.35±0.49 & 11.31±1.27 & 29.83±0.35 & 58.17±0.25 & \thibest{53.35±0.15} & 24.92±0.37 & 4.98±0.49 & 18.19±1.94 & 43.03±0.49 & 71.81±0.24 & \thibest{68.70±0.13} & 33.66±0.58 & 7.33±0.07 & 21.50±0.12 \\
\textbf{MoleRec}~\citep{yang2023molerec} & 3.16±0.70 & 12.07±0.42 & 29.62±0.23 & 57.60±0.07 & 52.56±0.33 & 25.61±0.03 & 4.72±0.69 & 19.37±0.64 & 42.49±0.24 & 71.06±0.02 & 67.99±0.28 & 34.41±0.16 & 6.83±0.12 & 21.55±0.60 \\
\textbf{RAREMed}~\citep{zhao2024leave} & 3.35±0.77 & 14.91±0.47 & 31.72±0.41 & \thibest{59.18±0.03} & \best{54.49±0.01} & 27.29±0.40 & 5.16±1.02 & 23.28±0.75 & 45.05±0.56 & \thibest{72.47±0.04} & \best{69.72±0.02} & 36.49±0.58 & 6.39±0.10 & 20.90±0.20 \\
\midrule
\textbf{LEADER}~\citep{liu2024large} & 5.49±0.63 & 15.61±0.75 & 26.14±0.51 & 56.15±0.43 & 50.75±0.14 & 25.80±0.25 & 7.86±0.70 & 24.23±1.10 & 38.25±0.65 & 69.77±0.41 & 66.47±0.12 & 35.03±0.30 & 8.28±0.32 & 16.66±0.14 \\
\textbf{FLAME}~\citep{fan2025fine} & \thibest{8.11±0.09} & \thibest{20.91±0.04} & \thibest{32.68±0.13} & 58.85±0.04 & 52.31±0.03 & \thibest{30.12±0.07} & \thibest{12.72±0.09} & \thibest{32.27±0.05} & \thibest{47.03±0.19} & 72.35±0.04 & 67.84±0.03 & \thibest{41.09±0.07} & 7.64±0.04 & 21.75±0.03 \\
\textbf{GPT-4o}~\citep{achiam2023gpt} & 0.85±0.28 & 6.66±1.42 & 14.68±2.17 & 36.01±2.73 & 24.12±2.21 & 14.58±1.65 & 1.88±0.57 & 12.31±2.45 & 24.87±3.18 & 51.66±3.11 & 36.72±2.61 & 22.68±2.33 & 6.02±0.02 & 28.03±0.34 \\
\textbf{GPT-5-mini}~\citep{openai2025gpt} & 3.50±0.49 & 11.29±0.32 & 11.78±0.31 & 27.30±0.29 & 23.04±0.22 & 13.50±0.10 & 6.50±0.76 & 18.89±0.53 & 19.48±0.46 & 39.52±0.31 & 36.44±0.22 & 21.10±0.10 & 11.60±0.03 & 11.37±0.16 \\
\textbf{GPT-5}~\citep{openai2025gpt} & 5.74±0.55 & 13.67±0.03 & 15.88±0.19 & 33.49±0.22 & 29.58±0.11 & 17.19±0.07 & 9.73±0.93 & 22.46±0.04 & 24.94±0.30 & 46.97±0.26 & 44.75±0.11 & 26.02±0.07 & 9.24±0.07 & 13.41±0.07 \\
\midrule
\multirow{2}[0]{*}{\method} & \best{13.27±0.97} & \best{23.05±0.45} & \best{33.40±0.49} & \best{60.04±0.24} & \secbest{53.53±0.10} & \best{32.43±0.23} & \best{20.61±1.47} & \best{35.67±0.71} & \best{48.26±0.53} & \best{73.48±0.21} & \secbest{68.85±0.08} & \best{44.51±0.41} & 
8.29±0.46 &
23.22±0.21 \\
         & (+ 30.17\%) & (+ 4.05\%) & (+ 1.86\%) & (+ 0.94\%) & (- 1.76\%) & (+ 4.07\%) & (+ 30.88\%) & (+ 3.16\%) & (+ 1.57\%) & (+ 0.60\%) & (- 1.25\%) & (+ 4.18\%) & (GT = 8.26) & (GT = 19.58) \\
        \bottomrule
    \end{tabular}
}
    \scalebox{0.78}{
        \begin{tabular}{c|cccc|cc|cccc|cc|cc}
        \toprule
        \multirow{2}[3]{*}{\textbf{\mimicfour}} & 
        \multicolumn{4}{c|}{\textbf{Group Jaccard}} & \multirow{2}[1]{*}{\textbf{\makecell{Micro\\Jaccard}}} & 
        \multirow{2}[1]{*}{\textbf{\makecell{Macro\\Jaccard}}} & \multicolumn{4}{c|}{\textbf{Group F1}} &
        \multirow{2}[1]{*}{\textbf{\makecell{Micro\\F1}}} & \multirow{2}[1]{*}{\textbf{\makecell{Macro\\F1}}} & 
        \multirow{2}[1]{*}{\textbf{\makecell{DDI\\Ratio}}} & 
        \multirow{2}[1]{*}{\textbf{\makecell{\# Predicted\\Medications}}} \\
        \cmidrule(lr){2-5}\cmidrule(lr){8-11} 
        &  \textbf{1 (Rare)} & \textbf{2} & \textbf{3} & \textbf{4 (Freq.)} & & & \textbf{1 (Rare)} & \textbf{2} & \textbf{3} & \textbf{4 (Freq.)} & &  \\
        \midrule \midrule
\textbf{Logistic Reg.} & 1.57±0.00 & 14.41±0.00 & 21.40±0.00 & 40.38±0.00 & 39.89±0.00 & 19.15±0.00 & 2.70±0.00 & 22.20±0.00 & 33.50±0.00 & 55.70±0.00 & 54.39±0.00 & 28.10±0.00 & 7.30±0.00 & 8.72±0.00 \\
\textbf{RETAIN}~\citep{choi2016retain} & 0.48±0.06 & 6.44±0.61 & 16.22±0.32 & 39.92±0.10 & 40.67±0.05 & 17.81±3.06 & 0.85±0.10 & 10.30±0.96 & 25.43±0.45 & 55.11±0.12 & 56.11±0.07 & 22.56±0.40 & 8.43±0.11 & 14.21±0.11 \\
\textbf{LEAP}~\citep{zhang2017leap} & 0.83±0.18 & 7.20±0.46 & 10.37±0.30 & 30.99±0.38 & 35.05±0.37 & 12.16±0.18 & 1.57±0.32 & 12.03±0.61 & 17.26±0.38 & 43.82±0.44 & 49.41±0.45 & 18.39±0.23 & 6.36±0.15 & 7.12±0.11 \\
\textbf{GAMENet}~\citep{shang2019gamenet} & 1.33±0.05 & 8.42±0.08 & 16.00±0.77 & 40.73±0.11 & 42.60±0.15 & 16.37±0.19 & 2.25±0.15 & 12.48±0.20 & 25.05±1.06 & 55.56±0.13 & 57.35±0.16 & 23.48±0.32 & 6.80±0.07 & 10.60±0.03 \\
\textbf{SafeDrug}~\citep{yang2021safedrug} & 1.38±0.57 & 9.93±0.85 & 19.58±0.61 & 42.76±0.17 & 43.37±0.09 & 18.13±0.18 & 2.31±0.91 & 14.30±1.09 & 30.41±0.73 & 58.16±0.23 & 58.31±0.10 & 25.90±0.14 & 6.35±0.19 & 12.58±0.17 \\
\textbf{COGNet}~\citep{wu2022conditional} & \thibest{7.72±0.69} & \secbest{24.05±0.46} & \best{28.25±0.09} & \thibest{45.91±0.11} & 44.07±0.19 & \thibest{26.17±0.32} & \thibest{12.96±1.10} & \secbest{36.60±0.51} & \secbest{42.75±0.10} & \thibest{61.60±0.13} & 58.48±0.26 & \secbest{38.06±0.46} & 8.03±0.19 & 19.05±0.14 \\
\textbf{MICRON}~\citep{yang2021change} & 0.13±0.09 & 7.87±0.97 & 19.66±0.23 & 43.37±0.19 & 44.09±0.03 & 15.92±0.19 & 0.25±0.18 & 12.21±1.57 & 30.89±0.42 & 59.01±0.25 & 59.12±0.04 & 23.24±0.34 & 6.78±0.03 & 12.54±0.24 \\
\textbf{MoleRec}~\citep{yang2023molerec} & 2.61±0.60 & 12.15±0.27 & 21.43±0.54 & 43.37±0.37 & 43.95±0.21 & 19.61±0.29 & 4.11±0.97 & 17.93±0.63 & 32.89±0.76 & 58.82±0.43 & 59.03±0.16 & 28.04±0.49 & 6.54±0.14 & 12.98±0.27 \\
\textbf{RAREMed}~\citep{zhao2024leave} & 1.50±0.30 & 16.17±0.98 & 24.27±0.54 & 45.58±0.16 & \secbest{45.72±0.08} & 21.54±0.49 & 2.60±0.52 & 24.22±1.48 & 37.01±0.80 & 61.16±0.18 & \secbest{60.64±0.06} & 30.78±0.74 & 14.51±10.77 & 12.89±0.21 \\
\midrule
\textbf{LEADER}~\citep{liu2024large} & 1.52±0.45 & 17.10±1.20 & 20.86±0.50 & 41.63±0.99 & 42.67±2.69 & 22.68±3.99 & 2.67±0.85 & 25.87±1.78 & 32.27±0.84 & 56.96±1.12 & 57.36±2.97 & 33.00±5.78 & 7.45±0.37 & 10.30±2.67 \\
\textbf{FLAME}~\citep{fan2025fine} & \secbest{10.17±0.14} & \thibest{22.15±0.09} & \thibest{27.65±0.35} & \best{46.30±0.05} & \thibest{44.70±0.14} & \secbest{26.30±0.13} & \secbest{16.01±0.28} & \thibest{33.40±0.13} & \thibest{41.37±0.50} & \best{61.86±0.03} & \thibest{59.43±0.19} & \thibest{37.80±0.19} & 7.50±0.11 & 11.58±0.38 \\
\textbf{GPT-4o}~\citep{achiam2023gpt} & 0.83±0.13 & 4.21±0.64 & 9.41±1.38 & 25.11±2.25 & 17.84±1.90 & 9.74±1.08 & 1.62±0.25 & 7.97±1.16 & 16.84±2.21 & 39.09±2.89 & 27.84±2.49 & 16.14±1.61 & 5.64±0.04 & 20.44±0.50 \\
\textbf{GPT-5-mini}~\citep{openai2025gpt} & 0.83±0.44 & 2.34±0.46 & 3.96±0.35 & 13.30±0.40 & 11.44±0.29 & 5.04±0.40 & 1.59±0.81 & 4.46±0.87 & 7.50±0.65 & 22.09±0.57 & 19.41±0.43 & 8.79±0.71 & 13.75±0.06 & 9.77±0.03 \\
\textbf{GPT-5}~\citep{openai2025gpt} & 4.28±0.13 & 9.85±0.36 & 13.70±0.16 & 24.60±0.03 & 22.49±0.03 & 12.96±0.03 & 7.48±0.21 & 16.95±0.53 & 22.93±0.23 & 37.01±0.03 & 35.05±0.04 & 20.87±0.03 & 11.29±0.02 & 10.87±0.02 \\
\midrule
\multirow{2}[0]{*}{\method} & \best{12.50±0.65} & \best{25.78±0.33} & \secbest{28.20±0.58} & \secbest{46.10±0.49} & \best{46.45±0.05} & \best{27.89±0.34} & \best{20.04±0.93} & \best{38.99±0.35} & \best{42.90±0.70} & \secbest{61.81±0.51} & \best{61.51±0.03} & \best{40.59±0.44} & 7.71±0.28 & 13.87±0.14 \\
         & (+ 22.84\%) & (+ 7.20\%) & (- 0.17\%) & (- 0.42\%) & (+ 1.59\%) & (+ 6.04\%) & (+ 25.20\%) & (+ 6.53\%) & (+ 0.36\%) & (- 0.08\%) & (+ 1.43\%) & (+ 6.66\%) & (GT = 7.89) & (GT = 11.46) \\
        \bottomrule
    \end{tabular}}
    \label{tab:main_result}
\end{table*}

\section{Experiments}
\label{sec:exps}
In this section, we review experiments for evaluating the effectiveness of our proposed framework, \method.

\subsection{Experimental Setup}
\label{sec:expsetup}
\smallsection{Datasets.}
We used two EHR datasets, \mimicthree~\citep{johnson2016mimic} and \mimicfour~\citep{johnson2023mimic}, that contain patient records of ICU patients.
For fair comparison, we did not incorporate any method-specific additional information, such as clinical notes or discharge summaries.
We also used drug-drug interaction (DDI) data from the TWOSIDES database~\citep{tatonetti2012data}.
We followed the same data pre-processing and dataset split protocol as in previous studies~\citep{yang2021safedrug,shang2019gamenet,wu2022conditional}, dividing all patient records into training, validation, and test sets with a ratio of 4:1:1.
We provide the statistics of the datasets in the online appendix~\citep{code}.

\smallsection{Evaluation.}
We evaluate the effectiveness of \method primarily using Jaccard Similarity ($\jaccard$) and F1 Score ($\fone$), widely adopted metrics in medication recommendation.
Specifically, we employ the following evaluation strategies:
\begin{itemize}[leftmargin=*, topsep=1pt, itemsep=1pt]
\item \textbf{Micro-Averaged (Micro):}
Aggregates predictions across all medications to compute the metrics. This strategy inherently assigns greater weight to frequently prescribed medications.
\item \textbf{Macro-Averaged (Macro):}
Computes the metrics for each medication individually and then averages them across all medications. This strategy ensures equal contribution from every medication regardless of prescription frequency.
\item \textbf{Group-Averaged (Group):}
Divides medications into groups according to their occurrence in the training set (28 per group in \mimicthree; 30, 30, 30, and 32 in \mimicfour), and reports group-wise averages of metrics.
This strategy reveals how performance varies across groups according to prescription frequencies.
\end{itemize}
These complementary strategies provide a more balanced assessment, allowing us to determine whether improvements on \raremeds are achieved without degrading performance for \freqmeds.
Additionally, we report the average number of predicted medications per admission and DDI ratios that evaluate the safety by measuring the proportion of predicted medication pairs known to have adverse interactions across all admissions.
We report the average and standard deviation of three trials. Refer to the online appendix~\citep{code} for details of metrics and evaluation strategies.

\smallsection{Baselines.} 
We compared \method with \numbaselines medication recommendation methods as follows:
Logistic Regression, RETAIN~\citep{choi2016retain}, LEAP \citep{zhang2017leap}, GAMENet~\citep{shang2019gamenet}, SafeDrug~\citep{yang2021safedrug}, MICRON~\citep{yang2021change}, COGNet~\citep{wu2022conditional}, MoleRec~\citep{yang2023molerec}, RAREMed~\citep{zhao2024leave}, LEADER~\citep{liu2024large}, 
FLAME~\citep{fan2025fine}, 
GPT-4o~\citep{achiam2023gpt}, GPT-5-mini, and GPT-5~\citep{openai2025gpt}.
Refer to the online appendix~\citep{code} for their details, including hyperparameter settings.

\smallsection{Details of \method.}
During counterfactual data generation, for each dataset, we generated $30$ counterfactual scenarios for each medication, including \raremeds.
For each generation, we selected $15$ associated diagnoses, $15$ procedures, and $10$ medications in descending order of relative risk with respect to the target medication $\targetmedi$.
To compute the relative risk, we considered only the training admissions.
Note that we used at most three visits in counterfactual data generation, and thus the maximum number of visits per scenario is capped at three.
Each counterfactual scenario targets a single medication and involves randomly sampling three patient candidates without prior prescriptions of that medication (i.e., $k=3$).
We utilized GPT-4o~\citep{achiam2023gpt} for counterfactual inference model $\llmcf$ with a temperature of $0.3$.
We present the statistics of the generated counterfactual data in the online appendix~\citep{code}.
In the subsequent steps, we utilized the LLaMA 3.1 8B Instruct model~\citep{grattafiori2024llama} as the instruction-tuned model $\llminst$ and later as the recommendation model $\llmrec$.
We provide additional implementation details for \method in the online appendix~\citep{code}.

\begin{table*}[t]
\vspace{-2mm}
    \caption{Ablation Study Results on \mimicthree and \mimicfour. 
    The best result for each column is highlighted in \best{yellow}. 
    The individual and combined contributions of counterfactual data (CF) and instruction tuning (IT) are confirmed.}
    \setlength{\tabcolsep}{1.0pt}
    \small
    \centering
    \scalebox{0.78}{
        \begin{tabular}{*{2}{>{\centering\arraybackslash}b{2em}|}cccc|cc|cccc|cc|cc}
        \toprule
        \multicolumn{2}{c|}{\textbf{\mimicthree}} & \multicolumn{4}{c|}{\textbf{Group Jaccard}} & \multirow{2}[1]{*}{\textbf{\makecell{Micro\\Jaccard}}} & \multirow{2}[1]{*}{\textbf{\makecell{Macro\\Jaccard}}} & \multicolumn{4}{c|}{\textbf{Group F1}} & \multirow{2}[1]{*}{\textbf{\makecell{Micro\\F1}}} & \multirow{2}[1]{*}{\textbf{\makecell{Macro\\F1}}} & 
        \multirow{2}[1]{*}{\textbf{\makecell{DDI\\Ratio}}} & 
        \multirow{2}[1]{*}{\textbf{\makecell{\# Predicted\\Medications}}} \\
        \cmidrule(lr){1-6}\cmidrule(lr){9-12} 
        $\mathbf{CF}$ & $\mathbf{IT}$ & \textbf{1 (Rare)} & \textbf{2} & \textbf{3} & \textbf{4 (Freq.)} & & & \textbf{1 (Rare)} & \textbf{2} & \textbf{3} & \textbf{4 (Freq.)} & &  \\
        \midrule \midrule
 &  & 10.21±0.93 & 20.73±1.91 & 30.04±0.94 & 59.03±0.55 & 52.63±0.49 & 29.97±0.86 & 15.42±1.18 & 32.30±2.76 & 44.01±1.21 & 72.58±0.51 & 68.11±0.42 & 41.08±1.12 & 8.34±0.28 & 21.95±0.51 \\
& \checkimg & 12.29±1.71 & 22.24±0.34 & 31.37±0.17 & 59.26±0.13 & 52.89±0.22 & 31.26±0.54 & 19.09±2.44 & 34.52±0.27 & 45.84±0.22 & 72.83±0.12 & 68.31±0.19 & 43.07±0.71 & 8.14±0.21 & 23.19±0.52 \\
\checkimg &  & 11.98±0.37 & 22.24±0.58 & 31.44±0.61 & 59.27±0.27 & 53.14±0.19 & 31.23±0.44 & 18.23±0.81 & 34.25±0.86 & 45.82±0.64 & 72.80±0.25 & 68.56±0.15 & 42.77±0.60 & 8.09±0.17 & 21.45±0.53 \\
\midrule
\checkimg & \checkimg & \best{13.27±0.97} & \best{23.05±0.45} & \best{33.40±0.49} & \best{60.04±0.24} & \best{53.53±0.10} & \best{32.43±0.23} & \best{20.61±1.47} & \best{35.67±0.71} & \best{48.26±0.53} & \best{73.48±0.21} & \best{68.85±0.08} & \best{44.51±0.41} & 8.29±0.46 & 23.22±0.21 \\
        \bottomrule
    \end{tabular}}
    \scalebox{0.78}{
        \begin{tabular}{*{2}{>{\centering\arraybackslash}b{2em}|}cccc|cc|cccc|cc|cc}
        \toprule
        \multicolumn{2}{c|}{\textbf{\mimicfour}} & \multicolumn{4}{c|}{\textbf{Group Jaccard}} & 
        \multirow{2}[1]{*}{\textbf{\makecell{Micro\\Jaccard}}} & \multirow{2}[1]{*}{\textbf{\makecell{Macro\\Jaccard}}} & 
        \multicolumn{4}{c|}{\textbf{Group F1}} & \multirow{2}[1]{*}{\textbf{\makecell{Micro\\F1}}} & 
        \multirow{2}[1]{*}{\textbf{\makecell{Macro\\F1}}} & 
        \multirow{2}[1]{*}{\textbf{\makecell{DDI\\Ratio}}} & 
        \multirow{2}[1]{*}{\textbf{\makecell{\# Predicted\\Medications}}} \\
        \cmidrule(lr){1-6}\cmidrule(lr){9-12} 
        $\mathbf{CF}$ & $\mathbf{IT}$ & \textbf{1 (Rare)} & \textbf{2} & \textbf{3} & \textbf{4 (Freq.)} & & & \textbf{1 (Rare)} & \textbf{2} & \textbf{3} & \textbf{4 (Freq.)} & &  \\
        \midrule \midrule
 &  & 8.77±0.48 & 22.43±0.09 & 25.53±0.50 & 44.93±0.33 & 46.04±0.19 & 25.32±0.18 & 14.38±0.71 & 34.13±0.15 & 39.31±0.64 & 60.45±0.26 & 61.17±0.19 & 36.91±0.23 & 7.44±0.16 & 12.48±0.08 \\
 & \checkimg & 12.43±0.68 & 24.25±0.46 & 26.78±0.58 & 44.99±0.13 & 45.93±0.21 & 26.87±0.25 & 19.77±0.99 & 36.90±0.52 & 41.02±0.78 & 60.55±0.13 & 60.99±0.21 & 39.23±0.35 & 7.49±0.14 & 13.61±0.15 \\
 \checkimg &  & 10.61±1.26 & 24.23±0.33 & 27.40±0.49 & 45.69±0.13 & 46.30±0.16 & 26.71±0.28 & 17.07±1.68 & 36.61±0.44 & 41.58±0.69 & 61.22±0.16 & 61.36±0.19 & 38.76±0.35 & 7.66±0.08 & 12.71±0.16 \\
 \midrule
 \checkimg & \checkimg & \best{12.50±0.65} & \best{25.78±0.33} & \best{28.20±0.58} & \best{46.10±0.49} & \best{46.45±0.05} & \best{27.89±0.34} & \best{20.04±0.93} & \best{38.99±0.35} & \best{42.90±0.70} & \best{61.81±0.51} & \best{61.51±0.03} & \best{40.59±0.44} & 7.71±0.28 & 13.87±0.14 \\
        \bottomrule
    \end{tabular}}
    \label{tab:ablation_result}
\end{table*}

\begin{table}[t]
    \centering
    \caption{Plausibility Evaluation Results for Counterfactual Data based on Human Expert and LLMs.}
    \label{tab:cf_plausibility}
    \scalebox{0.78}{
    \begin{tabular}{c|c|cccccc}
    \toprule
    \textbf{Evaluator} & \multirow{2}[1]{*}{\textbf{\makecell{Human\\Expert}}} & \multicolumn{6}{c}{\textbf{OpenAI gpt-\{model name\} (2025.08.07)}}  \\
    \cmidrule{3-8}
    \textbf{\& Metric} & & \textbf{4.1-nano} & \textbf{4.1-mini} & \textbf{4.1} & \textbf{5-nano} & \textbf{5-mini} & \textbf{5}  \\
    \midrule
    \midrule
    \textbf{Accuracy} & 0.45 & 0.50 & 0.59 & 0.56 & 0.51 & 0.59 & 0.61 \\
    \bottomrule
    \end{tabular}}
\end{table}

\subsection{Main Results (Table~\ref{tab:main_result})}
\label{sec:mainresult}
We present the comparison results of \method and \numbaselines baselines in Table~\ref{tab:main_result}.
Based on our empirical analysis, we highlight three key findings that provide insight into improving \raremeds prediction.

\smallsection{\method is effective, especially for \raremeds.}
\method outperforms all \numbaselines baselines across most evaluation metrics, demonstrating its general effectiveness in medication recommendation.

In terms of macro-averaged metrics, \method consistently achieves the best performance across both $\jaccard$ and $\fone$ regardless of the dataset. 
Numerically, \method outperforms the best-performing baselines (COGNet and FLAME) by relative improvements ranging from 4.07\% to 6.66\%, demonstrating its capacity to maintain robust performance across medications of varying prescription frequencies.
A more fine-grained analysis based on group-averaged scores further highlights the strengths of \method in handling rarely prescribed medications (\raremeds). 
Specifically, for Group~1, which includes the most rarely prescribed medications, \method significantly outperforms the strongest baseline (COGNet) by 30.17\% in terms of $\jaccard$ and 30.88\% in terms of $\fone$ on \mimicthree.
These substantial gains underscore the effectiveness of \method in improving the predictive performance for \raremeds by leveraging both (1) counterfactual data and (2) the LLM's capacity to compensate for limited training instances.
On the other hand, the performance gap narrows for groups including frequently prescribed medications (\freqmeds) and in micro-averaged metrics.
This is expected, since abundant training samples leave less room for improvement.
Nevertheless, \method performs comparably to the best-performing baseline (RAREMed) on both datasets in micro-averaged metrics and Group~4, which includes the most frequently prescribed medications.
In summary, \method achieves substantial gains in predicting \raremeds, as evidenced by large improvements in macro-averaged and group-averaged metrics, while preserving strong performance on \freqmeds under micro-averaged evaluation.

\begin{table*}[t]
\vspace{-1mm}
\centering
\caption{Clinical Plausibility Assessment of Generated Counterfactual Data. 
The `Related Original Elements' column lists diagnoses, procedures, or medications of the factual data related to the counterfactual additions. 
The `Counterfactual Additions \& Rationale' column presents each newly added element along with a brief justification of its clinical plausibility.
Additional counterfactual examples and their analyses are provided in the online appendix~\citep{code}.}
\small
\scalebox{0.82}{
\begin{tabular}{p{7cm}|p{16cm}}
\toprule
\multicolumn{1}{c|}{\textbf{Related Originial Elements}} & 
\multicolumn{1}{c}{\textbf{Counterfactual Additions \& Rationale (\textit{Target Medication},} \uls{Counterfactual Additions}\textbf{)}} \\
\midrule
\multicolumn{1}{c|}{\multirow{1}[15]{*}{\makecell{
\cfcyan{single live birth} \\
\cfpurple{secondary uterine inertia} \\
\cforange{gestational hypertension} \\ 
\cfbrown{chorioamnionitis} \\ 
\cfteal{opioid analgesic (N02A)} 
}}} & 
\textbf{\textit{Non-opioid analgesic class (N02B)}} provides postoperative pain control while limiting extra use of \cfteal{opioid analgesic (N02A)};
\uls{{Outcome of delivery, single liveborn}}, completes discharge coding for a \cfcyan{single live birth};  
\uls{{Abnormal fetal heart rate/rhythm}} precipitates \uls{\cfolive{{medical induction}}} in the context of \cfpurple{secondary uterine inertia}; 
\uls{\cfolive{{Medical induction of labor}}} addresses \uls{{fetal distress}} and corrects \cfpurple{uterine inertia};
\uls{{Low cervical cesarean section}} is required after failed \uls{\cfolive{medical induction}} with \cforange{gestational hypertension} and \cfbrown{chorioamnionitis};
\uls{{Osmotic laxative class (A06A)}} mitigates constipation caused by \cfteal{opioid analgesic (N02A)} and \uls{{cesarean surgery}} \\
\bottomrule
\end{tabular}}
\label{tab:anal_cf_data}
\end{table*}

\smallsection{Alignment is important for the effectiveness of LLMs.}
Our results indicate that effectively utilizing large language models (LLMs) for medication recommendation requires alignment between their clinical reasoning capabilities and the medication recommendation task.
Specifically, directly applying GPT models to medication recommendation in a zero-shot manner shows limited predictive performance.
Similarly, even with fine-tuning, LLMs lacking a proper adaptation step (e.g., LEADER) exhibit relatively poor predictive performance.
Notably, \method demonstrates substantial performance improvements over the LLM-based baselines by incorporating the proposed instruction-tuning step prior to fine-tuning.

\smallsection{Considering co-recommended medications is beneficial.}
Our results show that relationships among co-recommended medications, which have largely been overlooked in previous studies, may provide meaningful signals for improving the performance of \raremeds prediction.
Specifically, on \mimicthree, COGNet, which considers relationships between co-recommended medications through sequential decision making, shows strong macro-averaged and group-averaged performance. 
On \mimicfour, FLAME, which considers such relationships during its refinement step, also performs strongly on these metrics.
These results support the benefit of modeling relationships among co-recommended medications.

In addition, \method yields DDI ratios of $8.29$ on \mimicthree and $7.71$ on \mimicfour, which are close to those of the ground-truth prescriptions ($8.26$ and $7.89$, respectively).
These results indicate that \method does not materially increase the overall drug-drug interaction ratio (DDI) despite its substantial gain on \raremeds.\footnote{Medication recommendation is a high-stakes application, especially when models are considered for clinical decision support.
Note that DDI ratios alone do not fully characterize the clinical appropriateness of generated counterfactual samples or recommendations in broader decision-making settings.}

\vspace{-5pt}
\subsection{Ablation Study (Table~\ref{tab:ablation_result})}
\label{sec:ablation}
We conduct an ablation study to analyze contributions of counterfactual data and an instruction tuning and report results in Table~\ref{tab:ablation_result}.
Both instruction tuning and counterfactual data help to improve the predictive performance of \method. 
First, instruction tuning consistently improves predictive performance across all scores, demonstrating that the model may learn relevant clinical reasoning patterns from the instruction tuning.
That is, this suggests that the instruction-tuning step based on the medical record summarization task enhances \method's clinical reasoning capability to better capture patients' clinical context.
Additionally, counterfactual data significantly boosts performance for \raremed groups and macro-averaged metrics.
This result suggests that counterfactual data can mitigate data scarcity issues, pronounced for \raremeds.

\subsection{Plausibility Evaluation of Counterfactual Data with Human Expert Support (Table~\ref{tab:cf_plausibility})}
\label{sec:exp:plausibility}
We evaluate the plausibility of generated counterfactual data with a licensed human expert and multiple LLM evaluators on 100 samples (50 original and 50 counterfactual), each containing structured clinical elements (i.e., diagnoses, procedures, and medications). 
For counterfactual samples, these elements include modifications introduced during counterfactual data generation. 
Specifically, we conduct a blind discrimination test in which original and counterfactual samples are randomly sampled and mixed, and evaluators determine whether each sample is original or generated.
We use accuracy as the evaluation metric, where a value close to $0.5$ indicates that counterfactual data are as plausible as the original data and thus difficult to distinguish.

The human expert achieves an accuracy of $0.45$, indicating frequent misclassification between original and counterfactual samples. 
For LLM-based evaluation, we convert 1--5 plausibility scores~\citep{choi2025rag} into binary predictions using multiple thresholds; the resulting accuracies range from $0.50$ to $0.61$, showing that even strong LLMs struggle to reliably distinguish the two. 
Across both evaluations, the results indicate that neither the human expert nor strong LLMs consistently distinguish original data from generated counterfactual data. 
This suggests that the generated counterfactual samples preserve medical plausibility and contextual consistency to a degree that makes them difficult to separate from the original data.
We provide the detailed LLM evaluation setup and full results, including confusion matrices for all thresholds, in the online appendix~\citep{code}.

\subsection{Qualitative Analysis (Table~\ref{tab:anal_cf_data} and Table~\ref{tab:anal_rec_result})}
\label{sec:exp:qualitative}

\smallsection{Generated Counterfactual Data.}
In Table~\ref{tab:anal_cf_data}, we present a generated counterfactual example to illustrate and analyze the contextual plausibility of the newly introduced diagnoses, procedures, and medications (referred to as ‘counterfactual additions’ in the below).
In this example, a non-opioid analgesic (N02B) is added for postoperative pain control following obstetric surgery, and a counterfactual inference model $\llmcf$ constructs a coherent perinatal context by introducing delivery-related diagnoses and procedures.
This example suggests that $\llmcf$ can generate counterfactual additions that are relevant to the target medication, consistent with the original clinical context, and mutually coherent.

\smallsection{Recommendation Results.}
We qualitatively analyze the recommendation results presented in Table~\ref{tab:anal_rec_result}, comparing \method to COGNet and RAREMed, the strongest baselines in terms of micro- and macro-averaged performance, respectively.
In this example, only \method correctly predicts L02B and yields the fewest false negatives and false positives among the compared methods.
This example is consistent with our quantitative findings, showing that \method improves \raremeds prediction while remaining competitive on \freqmeds and producing fewer false positives.

To further support our qualitative findings, we provide additional generated counterfactual data and recommendation results, together with their detailed analyses, in the online appendix~\citep{code}.

\smallsection{Additional Experimental Results.}
The online appendix~\citep{code} provides a LoRA-rank sensitivity analysis showing that $\llmrec$ is robust to hyperparameter variations.
It also reports empirical training and inference runtimes as a practical reference.

\section{Conclusions and Future Work}
In this work, we address the critical challenge of accurately recommending rarely prescribed medications (\raremeds). 
Specifically, we focus on two inherent limitations in existing AI-based medication recommendation methods: (a) data scarcity and (b) overlooking relationships among co-recommended medications.

To mitigate these limitations, 
we propose a novel framework, \method, which leverages medical knowledge and clinical reasoning capabilities of large language models (LLMs). 
Specifically, we first develop a
counterfactual generation method to alleviate data scarcity issues, which are especially pronounced for \raremeds. 
We then integrate an LLM into our medication recommendation model to consider co-recommendation relationships.
Lastly, we introduce instruction tuning, aligning the LLM’s medical knowledge with the medication recommendation task
to better capture patients' context, including cases in which \raremeds should be prescribed.

\begin{table}[t]
\centering
\caption{Case-level Comparison of Recommended Medications.
The rarest medication is highlighted in \israre{orange}, and those recommended by all \uls{\textbf{three}} methods (omitted from \uls{\textbf{True Positive}} for brevity) are in \allright{gray}.
Additional results and analyses are provided in the online appendix~\citep{code}.}
\small
\scalebox{0.7}{
\begin{tabular}{c|c|c|c}
\toprule\multicolumn{1}{c|}{\textbf{Methods}} & \textbf{True Positive} & \textbf{False Negative} & \textbf{False Positive}  \\
\midrule
\textbf{Ground Truth} & \multicolumn{3}{c}{[\israre{L02B}, \allright{N03A}, \allright{A12B}, A12A, \allright{A02B}, \allright{N02A}, \allright{A12C}, \allright{N02B}, \allright{B01A}, \allright{A06A}]} \\
\cmidrule(lr){1-4}
RAREMed & None & [\israre{L02B}, A12A] & [C01B, N05A, N05B, A04A, A01A, B05C]  \\
COGNet & None & [\israre{L02B}, A12A] & [A11D, B03B, N05B, J01D, A07A, A01A, B05C] \\
\method (Ours) & [\israre{L02B}] & [A12A] & [J01E, J01D, A07A] \\
\bottomrule
\end{tabular}}
\label{tab:anal_rec_result}
\end{table}

Our experiments demonstrate that \method achieves up to a \improvement\% improvement in predictive performance for \raremeds compared to the best-performing method among \numbaselines baselines, without compromising performance for frequently prescribed medications. 
Additionally, our ablation study shows that both counterfactual data generation and instruction tuning are effective. 
Finally, plausibility evaluations by a licensed human expert and multiple LLM evaluators, together with qualitative analyses, provide additional evidence that the generated counterfactual samples preserve clinically coherent patterns and contextual consistency.

Future work can strengthen clinical validation through larger-scale evaluation with multiple clinicians and additional safety checks beyond DDI ratios, such as screening for drug-disease conflicts in generated counterfactual samples and final recommendations.

\vspace{1mm}
\smallsection{Acknowledgements.}
{\small This work was partly supported by the National Research Foundation of Korea (NRF) grant funded by the Korea government (MSIT) (No. RS-2024-00406985, 30\%).
This work was partly supported by Institute of Information \& Communications Technology Planning \& Evaluation (IITP) grant funded by the Korea government (MSIT) (No. RS-2024-00438638, EntireDB2AI: Foundations and Software for Comprehensive Deep Representation Learning and Prediction on Entire Relational Databases, 30\%)
(No. RS-2019-II190075, Artificial Intelligence Graduate School Program (KAIST), 10\%).
This work was partly supported by the Institute of Information \& Communications Technology Planning \& Evaluation (IITP)-ICT Creative Consilience Program grant funded by the Korea government (MSIT) (No. IITP-2026-RS-2020-II201819, 30\%).
}



\bibliographystyle{ACM-Reference-Format}
\bibliography{999refs}

\end{document}